\documentstyle[12pt]{article}
\def\IR{\relax{\rm I\kern-.18em R}}
\def\IN{\relax{\rm I\kern-.18em N}}
\def\cod{{\rm cod}}

\begin{document}

\title{Analysis of a three-component model phase diagram 
by Catastrophe Theory} 
%
%
\author{J. Gaite%
\thanks{Current address: Departamento de F\'{ \i}sica,
Facultad de Ciencias,
Universidad de Oviedo,
c/ Calvo Sotelo s.n.,
33007 OVIEDO.},
J. Margalef-Roig and S. Miret-Art\'es%
\thanks{Also: Max-Planck-Institut f\"{u}r Str\"{o}mungsforschung,
Bunsenstra{\ss}e 10, D-37073 G\"{o}ttingen, Germany.}
\\ 
Instituto de Matem\'{a}ticas y F\'{ \i}sica Fundamental \\
Serrano 123, 28006 Madrid, Spain.}  
\date{30 June 1997}

%
\maketitle

\begin{abstract}
We analyze the thermodynamical potential of a lattice gas model with three 
components and five parameters using the methods of Catastrophe Theory. We 
find the highest singularity, which has codimension five, and establish its 
transversality. Hence the corresponding seven-degree Landau potential,
the canonical form {\em Wigwam} or $A_6$, constitutes the adequate starting 
point to study the overall phase diagram of this model. 
\end{abstract}
%
%
\vskip 6cm
{\small FFUOV-97-5}\\{\small cond-mat/9707015}

\newpage

\global\parskip 4pt

\section{Introduction}
\label{Sec1}

The study of phase diagrams of complex systems is an important part of
Thermodynamics and applied physics. Phase diagrams can be constructed either
from experimental data or from theoretical models (for example, molecular
models). Among these, the wide class of lattice gas models is particularly
suitable for a mathematical analysis. Here we will focus our attention on the
lattice gas model for a system with three components which simulates, in
particular, a binary fluid mixture. A wide literature has already been devoted
to it from different points of view \cite[and references therein]{SST,griff2,
Mei,griff1}. We are mainly interested in attempting to present an overall 
analysis of its phase diagram, with particular attention to its highest 
multicritical point, that is, the one with the highest codimension in the 
five-dimensional parameter space that we consider.

In the mean field theory, the Gibbs potential is a function of the
concentration of two of the three components and depends on three
thermodynamical parameters, which can be taken as the temperature and the
chemical potentials of the two components, and on three molecular parameters.
The phase diagram deduced from this function is an accurate description of the
system, except close to the (multi)critical points, where fluctuations become
important and alter significantly the mean field theory predictions. For this
reason, the Gibbs potential has been the basis for determining the
overall phase diagram \cite{griff2,Mei}. The method used in \cite{griff2} 
establishes the qualitative features of the phase diagram, namely, 
the instability and (multi)critical (hyper)surfaces which divide its various 
regions, as well as some coexistence (hyper)surfaces. Those methods are 
considerably powerful but not sufficiently rigorous by mathematical standards. 
However, a  well established mathematical theory for the analysis of 
singularities of potentials and hence the associated phase diagrams does exist, 
namely, Catastrophe or Singularity Theory (CT).

Catastrophe theory has indeed been applied to the description of phase
transitions \cite{Schu, okada} and, in particular, of phase diagrams of complex
thermodynamical systems including fluid mixtures \cite{okada}. The philosophy 
behind these applications is however somewhat perverted: one starts with a 
system on which some knowledge of the phase diagram is available, perhaps 
its salient features, and surmises a polynomial (Landau) potential from among 
the variety supplied by the CT classification (called canonical forms), which 
is supposed to embody the properties of the thermodynamical potential of the 
system near the phase transition of interest. Usually, this potential is well 
analyzed in the mathematical literature and its other properties can be 
safely assigned to the physical system. In summary, this procedure is 
phenomenological in nature and amounts to fitting of phase diagrams. 
Although it utilizes CT,
it is only to take advantage of well studied potentials. However, this
way is, of course, less powerful and precise than the approach 
determining the polynomial potentials from a complete singularity analysis of 
the actual thermodynamical potential. It is this second approach the one we 
shall adopt here. A big advantage of our approach is that it is presented 
as a well defined algorithm leading to a systematic
way to analyze any general problem susceptible to be studied within the context
of CT.

Futhermore, this second alternative agrees with the methods of Ref.\ 
\cite{griff1, griff2}, which in fact approach those of CT from the point of 
view of classical thermodynamics. (See \cite{griff1} in regard to the 
convergence of both techniques).
We intend to take the best of both worlds for the problem in hand: to draw
intuition from thermodynamical methods and mathematical soundness from the
theorems of CT. In particular, we emphasize the study of transversality of the
actual thermodynamical potentials  which guarantees that those simple forms 
(polynomial potentials or canonical forms)  represent  indeed 
up to a diffeomorphism the original thermodynamical potential. We shall give a
brief account of the CT algorithm utilized throughout the paper together with 
an Appendix for more mathematical details.

Our results essentially agree with and support those in \cite{griff2}. However,
we hinge less on the visualization of the phase diagram and more in the 
classification of its singularities, relying for the construction of the phase 
diagram on the straightforward  method of gluing patches, 
each described by a standard canonical form for which the phase diagram can be 
found in the literature. Besides, we clearly establish the possibility of Landau
potentials in two variables, that is, of corank-2 canonical forms, 
for the system with three components. This possibility was dismissed in 
\cite{griff2}. Nevertheless, this case is sufficiently complex on its own 
to postpone it for future work. 

This work is organized as follows: In section 2 we introduce the fundamental 
concepts of CT, the more thechnical points of which are left for an appendix. 
In section 3 we describe the thermodynamical potential to be analyzed, 
give its physical interpretation and discuss general stability questions which
help connect usual concepts in Thermodynamics with those in Catastrophe Theory. 
In section 4 we apply the CT program to the potential previously introduced, 
reducing it first to a one-variable potential. In section 5 we show a solution
with the highest codimension and its transversality, thereby concluding that 
it is the {\em Wigwam} catastrophe. We also study in this section a peculiar 
singularity of lower codimension, associated to the physics of critical azeotropy. 
The last section is devoted to a discussion
of the previous results and of the structure of the phase diagram entailed by 
them.

\section{Generalities about Catastrophe Theory}

In this section we are going to review very briefly the main concepts of CT.
The reader desirous of more technical details is referred to Refs.\
\cite{poston,gilmore}. As is well known, 
CT deals with the singularities of smooth real-valued functions.
The nature of these singularities is revealed by perturbing those functions.
If as a result of a perturbation the qualitative properties of the function
remain unaffected we will say that this function is stable or structurally
stable. In other words, a function is said locally stable at a given point if 
there is a smooth change of coordinates so that the new function has the same 
structure as the old function. If a function is stable at all points then we
will say that this function is globally stable.
In a more precise way, a given function is a perturbation of another one at a
given point if the distance between both functions is arbitrary small. The 
concept of distance leads us to Topology. We can define the Taylor-series
topology in the space $\IR^d$ where $d$ gives the number  of the Taylor series
coefficients. Thus the $k$-jet of a given function at a given point is the 
Taylor series truncated beyond terms of degree $k$. Several definitions of 
distances can be given and all of these are expressed in terms of the $k$-jets 
of each function.

Now the next important question is what information is lost when we 
truncate the Taylor series of a function around a given point, namely, 
the problem of determinacy. It consists of determining whether a function 
can be truncated and if so, for what value of the degree of the Taylor 
expansion it can be truncated without any loss of substantial information. 
Furthermore, to determine the most general
family of functions of the smallest dimension $d$ which contains the original
function is called the problem of unfolding. The unfolding dimension is the
number of parameters describing a general perturbation and the minimum number 
to describe it is called the codimension. When all the unfolding terms go to
zero, the remainder of the universal unfolding is called the {\em germ} of the
canonical form.

The next step is to introduce the concept of transversality as a means to 
study structure stability and genericity. This concept was originally 
introduced by R. Thom \cite{thom} and, in general, is not widely used to 
classify physical phenomena in terms of elementary catastrophes. A property is
called generic if the subset for which the property is valid is open and dense 
in the original set. In other words, when a property is invariant under  a
perturbation, this property is called generic or structurally stable. 
The theorem of transversality shows that it is a 
generic property of functions to have only isolated {\it critical points}%
\footnote{We warn the reader not to confuse thermodynamical and mathematical 
terminology. In thermodynamical terms a critical point is just an equilibrium point
 and the 
word critical is reserved for a higher singularity.} and such functions are stable 
under perturbations. Two manifolds of $\IR^n$ intersect transversally if either
their intersection is empty or they intersect transversally at all their points
of intersection, that is, the direct sum of their tangent spaces at the point
has dimension $n$ or they span the tangent space $\IR^n$ at that point.

CT has usually not been applied in a rigorous way using all these concepts and
theorems needed for its correct implementation. We claim in this work that if 
this is done so, the procedure proposed by the theory is not by any means 
cumbersome and time consuming. The Catastrophe Program  proposed here 
provides a very useful and systematic way to examine with not very much effort
general behaviors of physical systems.

Let $F(x,\lambda)$ be a real function with state variables $x_1,\dots,x_n$ ($x 
\in \IR^n$) and control parameters $\lambda_1,\dots,\lambda_r$ ($\lambda
 \in \IR^r$);
that is, $F : \IR^{n+r} \rightarrow \IR $. We are to proceed as follows:\\
\begin{enumerate}

\item We pick ($x_0,\lambda_0$) such that $x_0$ is a {\em degenerate critical
point} of $F(x,\lambda)$ and we consider the unfolding
$f(x,\lambda)=F(x+x_0,\lambda+\lambda_0)-F(x_0,\lambda_0)$ and $h(x)=f(x,0)$. \\

\item One calculates the determinacy and codimension of $h$ from the $k$-jet of 
$h$ (see Appendix). 
Of course, if $h$ is $k$-determine then $h \sim j^k(h)$, that is, the
function $h$ is equal to $j^k(h)$ up to a change of coordinates and hence 
they are equivalent and have  qualitatively the same properties; 
therefore, cod$(h)=$ cod$(j^k(h))$.\\

\item One studies the $k$-transversality of $F$ and if this function is
$k$-tranversal we can affirm that $F$ and the canonical form of the unfolding
of $h$ are isomorphic and we can replace the original $F$ function for this 
canonical unfolding. If not, we can claim that the $F$ function is not 
susceptible to be studied by CT.\\
\end{enumerate}

\section{Description of the Gibbs potential}

According to Ref.\ \cite{griff2} a phenomenological model for a ternary mixture
is obtained by assuming the Gibbs potential in the form 
\begin{equation}
\label{GIBBS}
{\bar G} = N [a' yz + b' xz + c' xy + R\,T\,(x \ln x + y \ln y + z \ln z)],
\end{equation}
where $N=N_x + N_y + N_z$ gives the number of total moles and $N_x$,
$N_y$ and $N_z$ the moles of each component. The variables $x,y,z$ 
are the mole fractions defined by $x=N_x/N$, $y=N_y/N$ and $z=N_z/N$; 
and hence we have the constraint 
\begin{equation}
\label{CONS}
x + y + z = 1 
\end{equation} 
where $0 < x , y , z < 1$. Finally, $a', b'$
and $c'$ are phenomenological energy parameters. The energy
part is the most general quadratic term, given that $x + y + z = 1$. 
This model can be derived from the mean field theory of a lattice model 
Hamiltonian with variables taking three different states, representing the 
molecules of the three components \cite{KM}. Then $a',b'$ and $c'$ represent 
molecular interaction parameters.
Let us consider the Gibbs potential Eq.\ (\ref{GIBBS}) in a reduced form, 
dividing by $NRT$, and thus 
\begin{equation}
\label{RGIBBS}
G(x,y,z,a,b,c) = a\,yz + b\,xz +c\,xy + x \ln x + y \ln y + z \ln z
\end{equation}
where now the new parameters $a,b,c$ are defined with respect to 
the old ones $a',b',c'$ dividing them by $RT$.
The concentrations are supposed to be determined by some boundary 
conditions, such as the values of the chemical potentials of 
two components, say $\mu_x$ and $\mu_y$.  
The mean field theory prescription is then to minimize the {\it non-equilibrium
Gibbs potential} $G - \mu_x\,x - \mu_y\,y$ with respect to $x$ and $y$ 
to obtain the equilibrium conditions
$$
\frac{\partial G}{\partial x} = \mu_x,\\
\frac{\partial G}{\partial y} = \mu_y.
$$
They allow to solve for $x$ and $y$ as functions of $\mu_x,\mu_y$ and the 
parameters $a,b$ and $c$, provided that the Jacobian 
$\det\frac{\partial(\mu_x,\mu_y)}{\partial(x,y)} = \det{\partial_{ij}^2 G}$ 
is not zero. 

Thermodynamical stability further requires that the matrix 
${\partial_{ij}^2G}$ be positive definite. This property is called 
{\it convexity} 
and must hold for any thermodynamical potential, except on the instability 
hyper-surfaces, which are the simplest singularities we may encounter in a 
phase diagram. An instability can occur only near a phase transition, when two
equilibrium states, one unstable---hence unphysical---and the other 
meta-stable coalesce and disappear. In other words, a meta-stable state 
becomes unstable and, consequently, we speak of instability. This
is the kind of sudden change in the configuration of a system to which CT 
owes its name. In mathematical terms we say that the {\em critical point} 
(equilibrium state) is degenerate. As a consequence, one cannot solve for $x$ 
and $y$ as functions of $\mu_x,\mu_y$ or the solution is multi-valued, 
corresponding to the existence of 
various equilibrium states. The simplest instability occurs when only one 
eigenvalue of the stability matrix vanishes; that is, the instability only 
affects one variable. One is to focus on this variable, which is called 
{\it relevant}, to consider further singularities. Therefore, it is convenient
to transform the potential into a function of just the relevant variable by 
solving the equilibrium conditions for the other variables and
substituting for them. The standard thermodynamical procedure that performs 
this operation is the Legendre transform. In fact, this procedure can be used 
for potentials in other fields, whenever there is an underlying geometrical 
structure in the total space of variables, including state and control 
variables, called a {\em contact structure}. We address the reader interested 
in the general formulation to the literature \cite{ArnoldCS}. How the Legendre
transform is implemented in our case will be seen in the next section. 
Further singularities are studied afterwards with the one-variable potential. 
Next comes what can be called critical instability, followed by the 
tricritical point and so onwards. 

Several systems of interest are described by this Gibbs potential Eq.\ 
(\ref{GIBBS}): 
A ternary mixture at constant volumen, for example, a mixture of metals; 
a spin lattice where the molecules have spin one;   
a binary fluid mixture, where one of the three states represents a vacancy
instead of a new molecule and the corresponding concentration is associated to 
a variable total volumen. In the last case, the possible phases are
vapor, miscible liquid mixture and inmiscible liquid mixture. 
The convenient extensive variables are the specific volume $v$ and the relative 
concentration $\bar{x} = \frac{x}{x+y}$ of the two fluids and the intensive 
variables are the pressure and the chemical potential of one of the fluids. 
Moreover, the thermodynamical potential Eq.\ (\ref{GIBBS}) depends on $T,v$ and 
$\bar{x}$ and is therefore the Helmholtz potential $F(T,v,\bar{x}).$
This system is perhaps the most interesting
for applications, given the great amount of experimental data on 
binary fluid mixtures \cite{Row, McH-K}. However, the potential (\ref{GIBBS})
is not the most popular for fitting data; a related form which has 
similar dependence on the relative concentration of the two fluids but 
is of Van der Waals type for the volume is usually considered instead. We 
believe that this form, which is much more difficult to analyze, gives 
essentially the same qualitative behavior.

\section{Applying the CT program}

{}From Eqs.\ (\ref{RGIBBS}) and (\ref{CONS}), we have a function depending only on
two variables, $x$ and $y$, namely,
\begin{equation}
\label{HXY}
H(x,y,a,b,c) = a\, y (1-x-y) + b\, x (1-x-y) + c\,x y + x \ln x + y \ln y + 
(1-x-y) \ln (1-x-y)    .
\end{equation}

Now consider the function $H_y (x,y,a,b,c) - \mu_2$ (where the subindices
indicate derivatives with respect to the variable explicitely written
and $\mu_2 \equiv \mu_y$) and let 
the point $(x_0,y_0,a_0,b_0,c_0,\mu_2^0)$ be such that  $H_y(P_0)-
\mu_2^0 = 0$ and $H_{yy} (P_0) > 0$, where $P_0 =(x_0,y_0,a_0,b_0,c_0)$. 
The first condition is the  equilibrium condition for $y$ and 
the second one is required by stability in the $y$ direction. 
Then, by the Implicit Function Theorem, 
there exists a unique function $\psi (x,a,b,c,\mu_2)$ defined in a neighborhood
of $(x_0,a_0,b_0,c_0,\mu_2^0)$ with values in a neighborhood of $y_0$ 
such that $H_{yy} (x,y,a,b,c) > 0$, in these neighborhoods, and $H_y
(x,\psi(x,a,b,c,\mu_2),a,b,c) - \mu_2 = 0$ in the domain of $\psi$ and $\psi
(x_0,a_0,b_0,c_0,\mu_2^0) = y_0$.

By solving for $y$ as a function of $\mu_2$ and substituting into $H-\mu_2\,y
$ we have performed a Legendre transformation, effectively eliminating the 
variable $y$. Next we substract $\mu_1 x$ (where $\mu_1 \equiv \mu_x$) 
to obtain the function
\begin{equation}
\label{LX}
L(x,a,b,c,\mu_1,\mu_2) = H (x,\psi(x,a,b,c,\mu_2),a,b,c) - \mu_2\,  \psi
(x,a,b,c,\mu_2) - \mu_1 x,  
\end{equation} 
representing a one variable non-equilibrium Gibbs potential.
In order to have a function defined in a neighborhood of 
${\bar 0} = (0,0,0,0,0,0)$ we consider the new function
\begin{equation}
\label{LL}
L_1(x,a,b,c,\mu_1,\mu_2) =
L(x+x_0,a+a_0,b+b_0,c+c_0,\mu_1+\mu_1^0,\mu_2+\mu_2^0) - 
L(x_0,a_0,b_0,c_0,\mu_1^0,\mu_2^0)   .
\end{equation}
Then we have the following equations:
\begin{equation}
L_1({\bar 0}) = 0     \label{L10}
\end{equation}
and
\begin{eqnarray}
\lefteqn{L_{1,x} (x,a,b,c,\mu_1,\mu_2) = }  \nonumber \\ 
& & H_x\left[x+x_0,\psi(x+x_0,a+a_0,b+b_0,c+c_0,
\mu_2+\mu_2^0),a+a_0,b+b_0,c+c_0\right] \nonumber \\ 
& &  \mbox{}+ H_y \,\psi_x - (\mu_2+\mu_2^0)\, \psi_x - (\mu_1+\mu_1^0)
\end{eqnarray}
so that 
\begin{equation}
L_{1,x} ({\bar 0}) = H_x(P_0) -\mu_1^0   .  \label{L1x0}
\end{equation}
 
Suppose now that $H_x(P_0) - \mu_1^0= 0$ (the remaining equilibrium condition). 
Then $0$ would be called a {\it critical 
point} of $L_1(x,0,0,0,0,0)$ (in mathematical terminology). 
This function will be denoted by $g_1(x)$,
which is the {\em germ} to be studied. Of course, $g_1(0) = 0$ and $g_1' (0) = 0$ 
from (\ref{L10}) and (\ref{L1x0}). 
The two first derivatives of $g_1$ are
\begin{equation}
g_1' (x) = H_x(x+x_0,\psi(x+x_0,a_0,b_0,c_0,\mu_2^0),a_0,b_0,c_0) - \mu_1^0
\end{equation}
and
\begin{equation}
g_1'' (x) = H_{xx}(x+x_0,\psi(x+x_0,a_0,b_0,c_0,\mu_2^0),a_0,b_0,c_0) + 
H_{xy} (-) \psi_x (-)   .
\end{equation}
In particular 
\begin{equation}
g_1'' (0) = H_{xx} (P_0) + H_{xy}(P_0)\, \psi_x (x_0,a_0,b_0,c_0,\mu_2^0)
\end{equation}
with 
\begin{equation}
\psi_x (x_0,a_0,b_0,c_0,\mu_2^0) = \frac{- H_{xy} (P_0)}{H_{yy} (P_0)} ,
\end{equation}
as deduced from $H_y-\mu_2 = 0$ by taking the derivative with respect to $x$.
Suppose that the Hessian of $H$ is such that 
\begin{equation}
H_{xx} (P_0)\,H_{yy} (P_0) - H_{xy}^2 (P_0) = 0   ,
\end{equation}
then $g_1'' (0) = 0$ and $0$ is a degenerate critical point of 
$g_1$. Finally, we also assume that
$g_1''' (0) = g_1^{iv} (0) = g_1^v (0)
= g_1^{vi} (0) = 0$. Then we have imposed five conditions on $g_1$ 
altogether and we should be able to solve for 
$(x_0,y_0,a_0,b_0,c_0)$.
In this case, we say that we have reached the highest codimension. We will see
in the next section that there is indeed such a solution.

\section{Results}

\subsection{Highest singularity}

Now we look for a point which fulfills the 5 conditions for the highest singularity
mentioned above. We succesively have that 
\begin{equation}
y_0 = 1 - 2\,x_0,
\end{equation}
\begin{equation}
b_0 = x_0^{-1},
\end{equation}
\begin{equation}
a_0 = c_0 = \frac{1 + 2\, x_0}{8\, x_0\, (1- 2\, x_0)}
\end{equation}
and
\begin{equation}
36\, x_0^2 + 4\, x_0 - 1 = 0   .
\end{equation}
Thus from the last equation we have that $x_0 = \frac{\sqrt{10}-1}{18}\simeq 
0.1201265$. 
Moreover, $$g_1^{vii} (0) = 6\, x_0^{-2} \left( \psi_{5x} - \frac{1472}{27} \, 
x_0^{-4}\right) \neq 0,$$ where $$\psi_{5x} = \frac{256}{162}\, x_0^{-2} 
\left[ \frac{157}{10}\, x_0^{-2} - 17 (1-2x_0)^2
+ 5\, \frac{3x_0^2 -4x_0 +1}{x_0^2 (1-2x_0)^2}\right].$$ 

Now we apply results of Singularity Theory \cite{thom,ww}:

\begin{itemize}

\item The 7-jet of $g_1$ is $j^7 (g_1) = \frac{1}{7!}\, g_1^{vii}(0) \, x^7$ 
with $g_1^{vii} (0) \neq 0$. 

\item the essence of $g_1$ respect the identity is 7 and therefore 
$\sigma(g_1) \geq 7$, where $\sigma (g_1)$ is the determinacy of $g_1$
(see Appendix for the definition of this concept).

\item The codimension of $j^7 (g_1)$ is ${\rm cod}(j^7 (g_1)) = 
{\rm dim~vect}( \langle x\rangle  / \langle x^6 \rangle) =  5 $ 
(see Appendix) and $\sigma (j^7 (g_1)) \leq 7$. Thus $j^7 (g_1)$ is
7-determinate, $j^7 (g_1) \sim g_1$ and $g_1$ is 7-determinate. Moreover,
${\rm cod} (g_1) = 5$ and $\sigma (g_1) = 7$. A basis of this 
quotient vector space is 
given by the set $\{ [x],...,[x^5] \}$. Then ${\bar g}_1 (x,\lambda_1,...,
\lambda_5) = g_1(x) + \lambda_1 x + \lambda_2 x^2 + ... + \lambda_5 x^5$ is a 
{\it canonical unfolding} $k$-transversal of the {\em germ} $g_1$ for every $k > 0$;
in particular, for $k=7$.  Finally $g_1 \sim x^7$.

\item The $L_1$ function is an unfolding 7-transversal of $g_1$ because 
one can prove
(see Appendix)
\begin{equation}
\label{TRANS}
\langle x\rangle  = \langle x^6\rangle  + V_{L_1} + \langle x\rangle ^{7+1},
\end{equation}
where $V_{L_1}$ is the real linear space generated by
$$\{L_{1,a} (x,{\bar 0}) - L_{1,a} (0,{\bar 0}),\dots,L_{1,{\mu_2}} (x,{\bar 0}
) - L_{1,{\mu_2}} (0,{\bar 0}) \},$$ where the subindices $a,\dots,\mu_2$
denote derivatives with respect to the corresponding parameters. 
Note that Eq. (\ref{TRANS}) has the following expression
\begin{eqnarray}
\langle x \rangle  =  \langle x^6 \rangle  + 
\left\{ x \,(-1+2x_0) 
+ x^2 \,\frac{2}{3} x_0^{-1}\, (1-3x_0) + 
x^3 \left(\frac{2}{3} x_0^{-1} + \frac{4}{9} x_0^{-2} (-1 + 3x_0)\right)+ \right. 
\nonumber \\ 
 x^4 \left(- x_0^{-2} \frac{8}{9} + (3x_0-1)\, x_0^{-3} \frac{-352}{1080}\right) +
x^5 \left(\psi_{5x} (3x_0-1) \frac{1}{5!} + \frac{1}{4!} \frac{352}{45} x_0^{-3} 
+ \frac{16}{27} x_0^{-3}\right), \nonumber \\
- \frac{1}{3} x^2 + x^3 \frac{2}{9} x_0^{-1} + x^4 \left( - \frac{4}{9}
x_0^{-2} + \frac{1}{24} \frac{352}{45} x_0^{-2}\right) + x^5 \left(\frac{1}{24}
\frac{352}{45} x_0^{-3} - x_0 \frac{1}{5!} \psi_{5x}\right), \nonumber \\
x (1-2x_0) - \frac{2}{3} x^2 + x^3 \frac{-2}{9} x_0^{-1} +
x^4 \frac{1}{24} \frac{128}{45} x_0^{-2} + x^5 \frac{1}{5!} \left(x_0\, \psi_{5x}
- \frac{352}{9} x_0^{-3}\right), \nonumber \\ \left.
- x, \,x^2 \,\frac{2}{3} x_0^{-1} - x^3 \,\frac{4}{9} x_0^{-2} + \frac{1}{4!} x^4\,
 x_0^{-3}\, \frac{352}{45} + \frac{1}{5!} x^5\, (- \psi_{5x}) \right\},\nonumber
\end{eqnarray}
\begin{equation} \label{TRANS1}
\end{equation}
\noindent where the Taylor expansions in  $V_{L_1}$  have been truncated at
$x^5$ because the terms of degree sixth and higher are already included in 
$\langle x^6 \rangle$. The resolution of Eq.\ (\ref{TRANS1}) amounts to prove 
that a generic fith-degree polynomial with no independent term can 
be generated as a linear combination of the five polynomials between the curly 
brackets. Hence it implies the resolution of a linear system whose determinant 
is  $x_0\,( - 58944\, x_0^2 + 729 \, \psi_{5x}) \neq 0$.

Finally, this equality, Eq.\ (\ref{TRANS1}), holds and consequently 
${\bar g}_1$ and $L_1$ are isomorphic as unfoldings and $L_1$ can be 
qualitatively studied by the polynomial
$$
x^7 + \lambda_1\, x + \lambda_2\, x^2 + \lambda_3\, x^3 + \lambda_4\, x^4 + 
\lambda_5\, x^5,
$$
which in the terminology of CT corresponds to the {\it Wigwam} or $A_6$ 
catastrophe.
 
\end{itemize}

As a result of this analysis we can affirm that there are three changes of
coordinates $\varphi_1$, $\varphi_2$ and $\varphi_3$ and a perturbation
$\varepsilon$ of parameters such that
\begin{equation}
\label{LLB}
L_1(x,a,b,c,\mu_1,\mu_2) = u^7 + \lambda_1 u + \lambda u^2 + \cdots +
\lambda u^5 + \varepsilon (a,b,c,\mu_1,\mu_2)
\end{equation}
with
$$
u = \varphi_3\, p_1\, \varphi_1 (x,a,b,c,\mu_1,\mu_2)  ,
$$
$$
\varphi_2 (a,b,c,\mu_1,\mu_2) = (\lambda_1,\dots,\lambda_5)  .
$$
where $p_1$ means the {\it first projection}, $p_1 : \IR^{1+5}
\rightarrow \IR $, that is, $p_1 (x_1,\dots,x_6) = x_1$.

Moreover, the bifurcation set of Eq.\ (\ref{LL}) and that corresponding to
\begin{equation}
\label{WIG}
u^7 + \lambda_1 u + \lambda u^2 + \cdots + \lambda u^5 
\end{equation}
are diffeomorphic and we rather work with Eq.\ (\ref{WIG}) due to its 
simplicity. The equilibrium manifold in $(u,\lambda_1,\dots,\lambda_5)$ 
is obtained from Eq.\ (\ref{WIG}) by equating the first derivative to zero
$$
7 u^6 + \lambda_1 + 2 \lambda_2\, u + 3 \lambda_3\, u^2
+ 4 \lambda_4\, u^3 + 5 \lambda_5\, u^4 = 0
$$
and, furthermore, instability occurs if the second derivative also vanishes,  
$$
42 u^5 + 2 \lambda_2 + 6 \lambda_3\, u + 12 \lambda_4\, u^2
+ 20 \lambda_5\, u^3 = 0 .
$$
Now the bifurcation set is obtained by a projection onto the parameter space, 
that is, by eliminating the variable $u$ in this system of two equations. 

\subsection{Critical azeotropy as a singularity}

In the process of solving the equations that lead to the highest singularity one 
goes through singularities of lower codimension. They have no particular interest 
by themselves except in one case, which we proceed to describe. 

Thus we analyze now the five conditions one by one.  
The equilibrium conditions allow one to express $x_0$ and $y_0$ as functions of
the parameters, resulting in $y_0 =\psi(x_0,a_0,b_0,c_0,\mu_2^0)$, already used to 
define $g_1(x)$, and an equation for $x_0$ derived from equating (\ref{L1x0}) 
to zero. However, we prefer to keep $x_0$ and $y_0$ in the equations to follow, 
for it is simpler, understanding that they are to be substituted in the end. 
 
The first condition on $g_1(x)$ is $g_1'' (0) = 0$ or
\begin{equation}
\det H_{ij} = H_{xx} (P_0)\,H_{yy} (P_0) - H_{xy}^2 (P_0) = 0 .
\end{equation}
The Hessian matrix is 
\begin{equation}
\left(H_{ij}\right) = \left(\begin{array}{lr} \alpha_3 + \alpha_1 & \alpha_3\\
                            \alpha_3 & \alpha_3 + \alpha_2  \end{array}\right),
\label{Hessm}
\end{equation}
with $\alpha_3 = c-a-b + z^{-1}$, $\alpha_1 = a-c-b + x^{-1}$ y $\alpha_2 = b-a-c + 
y^{-1}$. We have reinstated $z=1-x-y$ for the sake of symmetry and we suppress 
the subindices zero relative to $P_0$ in the next equations. 
From Eq.\ (\ref{Hessm}), 
\begin{equation}
\det H_{ij} = \alpha_3\, \alpha_1 + \alpha_3\, \alpha_2 + \alpha_1\, \alpha_2=0,
\end{equation}
which is a quadratic equation on either $a,b,c$ or $x,y$.
The next condition $g_1'''(0) = 0$ is equivalent to
\begin{equation}
{1\over x^{2}}\,\alpha_2^{3} + {1\over y^{2}}\,\alpha_1^{3} - 
{1\over x^{2}}\,(\alpha_1+\alpha_2)^{3} = 0
\end{equation}
or
\begin{equation}
{1\over x^{2}}\,\alpha_3^{3} - {1\over y^{2}}\,(\alpha_3+\alpha_1)^{3} +
{1\over z^{2}}\,\alpha_1^{3} = 0
\end{equation}
or
\begin{equation}
-{1\over x^{2}}\,(\alpha_3+\alpha_2)^{3} + {1\over y^{2}}\,\alpha_3^{3} +
{1\over z^{2}}\,\alpha_2^{3} = 0.
\end{equation}
Each of these is an equation of third degree in $a,b,c$. 

Next we consider 
$g_1^{iv}(0) = 0$. Again this equation can be separated into three symmetric 
options. Each one imposes new conditions on $\alpha_i$ and hence on the elements
of the Hessian matrix. Furthermore, they can only be fulfilled if two 
new quantities vanish. In other words, the codimension increases by two units with
only one condition, a non-generic situation.
We choose the solution $\alpha_3=\alpha_1=0$ such that $H_{xx}(P_0) = 
H_{xy}(P_0) = 0$. There are two more solutions, obtainable by cyclic permutation
of the labels of the alphas. As a counterpart of the previous extra increase 
of the codimension by one unit, 
the next condition is identically fulfilled, $g_1^{v}(0) \equiv 0$. Finally, 
from $g_1^{vi}(0) = 0$ one obtains the solution quoted in the subsection above.

Let us compare the foregoing analysis with the one in the previous literature 
on this model \cite{KM,griff2}. This analysis goes as follows.
The first condition $g_1''(0) = 0$ or, equivalently, that the Hessian of $H$
be null has two types of solutions, namely, a simple solution,  
$\alpha_1^{-1}+\alpha_2^{-1}+\alpha_3^{-1}=0$, and a second
solution that requires that two of the $\alpha_i$ vanish simultaneously. 
The first type is more
generic but, unfortunately, does not lead to a high codimension singularity, 
since the equation $g_1^{iv}(0) = 0$ has no solution for it. The second
solution is the one we have considered in the previous subsection. 
It is called a symmetric solution 
because the additional condition for the Hessian matrix 
elements implies a symmetry in the phase diagram, namely, $a = c$. 

Alternatively, we can avoid making any choice on the type of solution until 
the last moment, namely, when we demand $g_1^{iv}(0) = 0$. Then this condition 
implies by itself $g_1^{v}(0) = 0$. In other words, the singularities given by 
these two conditions are 
inextricably linked: the first one entails the second one. Moreover, 
the first condition led us to take $H_{xx}(P_0) = H_{xy}(P_0) = 0$ in 
addition to $\det H_{ij} = 0$. We may recall here that the vanishing of these 
two elements of the stability matrix for a binary
fluid mixture has a thermodynamical interpretation: It occurs when there is 
critical azeotropy \cite[pages 197--199]{Row}. Azeotropy is not a singularity
on its own but when it superposes on a critical point it enhances its singularity
producing a new one. We see what kind of singularity it is in our case, namely,
the one given by $g_1^{v}(0) = 0$---a tricritical point. At this moment, we are 
not able to say if this is just a peculiarity of the particular three-component
model we study or in fact constitutes a general feature.

\section{Discussion}

The solution with the highest codimension, namely, five, which has been found above must be isolated; that is,
it cannot belong to a continuos family of solutions. Nevertheless, there can
be a discrete set of solutions. In fact, we can obtain two other solutions
from this one by using the symmetry of the original potential $G(x,y,z,a,b,c)$
under simultaneous permutations of $(x,y,z)$ and $(a,b,c)$. The way in which 
they arise is clear in the last subsection. It is clear as well that there are no
further solutions with codimension five.

Some readers may be concerned by the fact that our solution is a Landau 
potential which is not bounded below. This does not mean that there is no 
absolute minimum and the Landau potential is meaningless. We must remember 
that the Landau potential is a local
object and provides no information on the behavior of the thermodynamical 
potential far from the point that we have called $P_0$. This point corresponds
to a degenerate instability rather than to a multicritical point. To be 
precise, it arises as a tricritical point, with Landau potential $x^6$, 
merges with an unstable equilibrium state and disappears, for which the 
appropriate name is tricritical unstable point. Since $g_1^{vii} (0) \neq 0,$ 
there is no tetracritical point in the phase diagram. In this
we disagree with \cite{griff2}, where they assert to have three tetracritical 
points with $g_1^{vii} (0) = 0.$ We attribute this disagreement to a slip on 
their part, since the conditions they obtain for their tetracritical points are 
precisely the same five conditions we have for our tricritical unstable point 
and the codimension of a generic tetracritical point is six. Thus it seems 
that they are calling tetracritical points to what actually are tricritical 
unstable points.

The topology of the overall phase diagram is formed by gluing 3 patches 
corresponding each to the phase diagram of the {\em Wigwam} catastrophe, as 
given in the literature \cite{LNM}. The Taylor series expansions on which 
the CT is based are supposed to be valid in each patch. The coordinates on each 
patch are not related and therefore we can only obtain topological information.
Even the topological matching is not a trivial matter when we deal with 
high-dimensional spaces. Since the highest singularities 
in our analysis are essentially the same as those in \cite{griff2}, so is the 
phase diagram. The phase diagram is of great utility for experimentalists and 
we should like to obtain some more concrete information. We must notice that 
in a particular model, for example, a binary fluid mixture, the energy 
parameters are fixed and one can only tune the chemical potentials---one of 
which is to be interpreted as pressure in the binary mixture---and the 
temperature. Therefore, one is interested in three-dimensional 
sections of the overall phase diagram. In these sections it is not generic 
to have a tricritical point, (which has codimension four,) a fact well known 
to experimentalists. Generic and non generic sections of the phase diagram of 
the {\em Wigwam} catastrophe  are expected to cover all the possibilities.

Another solution of $\det H_{ij} = 0$ is, of course, that all the matrix
elements be null so the matrix has rank zero or co-rank 2. This means that one 
cannot use the Implicit Function Theorem to reduce to a function of one 
variable and one is to proceed with the CT program for a function of two 
variables. The next step is to analyze the 3-jet of this function, that is, 
the form given by the third derivatives. According to its signature, given by 
the sign of its discriminant, there are three possible cases: If it is 
negative, the canonical form is $x^3 - 3\,x\,y^2$, called the {\it elliptic 
umbilic} catastrophe; if it is positive the canonical
form is $x^3 + 3\,x\,y^2$, called the {\it hyperbolic umbilic} catastrophe; 
if it is null, the 3-form is degenerate, and one is to anlyze the
4-jet to determine the type of singularity, which may be the {\it parabolic 
umbilic} catastrophe or a type even more complex. The calculations driving at 
establishing the highest singularity in our five dimensional parameter space 
are complicated and the results for co-rank two shall be reported in the 
future \cite{gmm}. 

\section{Acknowledgements}

This work has been supported by DGICYT-Spain with Grants PB92-0302,
PB93-0454-C02-01 and PB95-0071. S.M.A. gratefully acknowledges the Alexander 
von Humboldt Foundation for a Fellowship.

\newpage

\appendix
\section{Summary of Catastrophe Theory}

In this Appendix we are going to present the main mathematical concepts
widely introduced in Refs.\ \cite{ww,lander} and necessary to follow the main 
steps developed in Section 5.

Let us consider real functions of class $\infty$ and defined in a neighbourhood
of $0 \in \IR^n$. We establish that two functions are equivalent if they
coincide in a neighbourhood of $0$. The classes we obtain are called {\em germs} of 
functions and the set of {\em germs} is denoted by $E(n)$. The operations $f+g$ and
$f\cdot g$ give to $E(n)$ the estructure of a ring and $M(n) = \{ f \in E(n) /
f(0) = 0 \}$ is a maximal ideal of this ring. Moreover, the operations 
$f+g$ and $\lambda \cdot f$ with $\lambda \in \IR$ give to $E(n)$ the structure of
a real vector space of dimension $\infty$.  The ideal $M(n)$ is generated by
$x_1,\dots,x_n$, that is, $M(n) = \{ f_1 x_1 + \cdots + f_n x_n \;/\, f_1,\dots,
f_n \in E(n) \}$. In general, if $f_1,\dots,f_n \in E(n)$, we designate by
$\langle f_1,\dots,f_n \rangle $ to the ideal generated by $f_1,\dots,f_n$,
that is, 
$$
\langle f_1,\dots,f_n \rangle = \{ f_1 g_1 + \cdots + f_n g_n \,\, /
\, g_1,\dots,g_n \in E(n) \}  .
$$
In particular, $M(n) = \langle x_1,\dots,x_n \rangle$.

It is possible to define powers of $M(n)$ as $M(n)^k$. It can be proven that
$M(n)^k$ is equal to the ideal of $E(n)$ generated by the monomials in $x_1,
\dots,x_n$ of degree $k$. In particular, for example, $\langle x,y \rangle ^2 = 
\langle x^2,xy,y^2 \rangle$. We have also that 
$M(n)^{k+1} = \{f \in E(n) / D^if(0) = 0,  i \leq k \}$ where with $D^i$
we mean the derivative of degree $i$.

In a similar way, we can define the {\em germs} of dipheomorphisms of class $\infty$
from $\IR^n$ in $\IR^n$ which transform $0$ in $0$. This set is denoted by 
$G(n)$.
We say that two {\em germs} are equivalent if there exists a change of coordinates 
$\varphi \in G(n)$ such that $f= g \,\varphi$ and it is denoted by $f \sim g$. 
When the $k$-jets ($k \in \IN$ )of two functions are equal we say that these
functions are $k$-equivalent  ($f \sim _k g$). 
A {\em germ} $f$ is $k$-determinate if for every {\em germ} $g$ such that both 
$k$-jets are equal we have that $f \sim g$. The determinacy of a {\em germ} $f$ is 
the smallest number $k \in \IN $ such that $f$ is $k$-determinate and it is called 
$\sigma (f)$. Therefore, we notice that :
\begin{description}

\item[] If f is $k$-determinate then $f \sim j^k f$ ,

\item[] If f is $k$-determinate and $f \sim _k g$ then $g$ is $k$-determinate ,

\item[] If f is $k$-determinate and $f \sim g$ then $g$ is $k$-determinate. 
\end{description}

The ideal of Jacobi of a {\em germ} $f$ is defined by
$$
\Delta (f) = \langle D_{x_1} f,\dots,D_{x_n} f \rangle
$$
where $D_{x_1},\dots,D_{x_n}$ are the partial derivatives with respect to the 
$x_1,\dots,x_n$ variables.  If $f \sim g$ then $ \Delta (f) \equiv \Delta (g)$.

Now let us assume that $f \in M(n)^2$. Then $\Delta (f) \subset M(n)$ and we can
speak of the quotient vector space  $M(n) / \Delta (f)$. The dimension of this 
vector space is called codimension of $f$, $\cod (f)$. It can be proven that if 
$f \in M(n)$
then  $\cod (f)$ is finite if and only if $\sigma (f)$ is finite and in this
case $\sigma (f) - 2 \leq \cod(f) $. Moreover, if $f \sim g$ then $\cod (f) =
\cod (g)$. Thus the codimension generally coincides with the number of conditions 
necessary to specify a function germ up to diffeomorphisms, which is the usual 
geometrical concept. 

If $g \in M(n+r)$ and $f \in M(n)^2$, we say that $g$ is a $r$-unfolding of $f$ if 
$g(x,0) = f(x)$ (with $r$ parameters which we define as $y_1,\dots,y_r$).
Now let us assume that $g \in M(n+r)$ is an unfolding of $f \in M(n)^2$ and $k
\in \IN$. We say that $g$ is $k$-transversal if 
$$
M(n) = \Delta (f) + M(n)^{k+1} + V_g
$$
where $V_g$ is the real vector subset of $M(n)$ generated by the
vectors $D_{y_1} g(x,0) - D_{y_1} g(0,0),\dots,D_{y_r} g(x,0) - D_{y_r} g(0,0)$.

Finally, the theorem of $k$-transversality for unfoldings can be stated as
follows: Let us consider $ f \in M(n)^2$ $k$-determinate and $g$ and $h$
two unfoldings of $f$ with $r$ parameters which are $k$-tranversal. 
Then $g$ and $h$ are isomorphic.


\newpage

\begin{thebibliography}{99}
%
\bibitem{SST} J.A. Schouten, A. ten Seldam and N.J. Trappeniers, Physica 
{\bf 73}, 556 (1974).
%
\bibitem{griff2} D. Furman, S. Dattagupta and R. B. Griffiths, Phys. Rev. {\bf
B15}, 441 (1977).
%
\bibitem{Mei} P.H.E. Meijer and M. Napi\'orkowski, J. Chem. Phys. {\bf 86}, 
5771 (1987).
%
\bibitem{griff1} R. B. Griffiths, Phys. Rev {\bf B12}, 345 (1975).
%
\bibitem{Schu} L.S. Schulman and M. Revzen, Collective Phenomena {\bf 1}, 43 
(1972).
%
\bibitem{okada} K. Okada, {\it Catastrophe Theory and Phase Transitions},
Scitec Publications, Zug (Switzerland), 1993. 
%
\bibitem{poston} T. Poston and I. N. Stewart, {\it Catastrophe Theory and Its
Applications}, Pitman, London, 1978.
%
\bibitem{gilmore} R. Gilmore, {\it Catastrophe Theory for Scientists and
Engineers}, Dover Publications, Inc., New York, 1981.
%
\bibitem{thom} R. Thom, {\it Stabilit\'e Structurelle et Morphog\'enese}, 
Benjamin, New york, 1972. 
%
\bibitem{KM} S. Krinsky and D. Mukamel, Phys. Rev {\bf B11}, 399 (1975).
%
\bibitem{ArnoldCS} V.I. Arnold, S.M. Gusein-Zade and A.N. Varchenko,
{\it Singularities of differentiable maps}, Birk\"auser, Basel 1985.
\\J.-G. Dubois and J.-P. Dufour, Ann. Inst. Henri Poincar\'e 
{\bf XXIX}, 1--50 (1978).
%
\bibitem{Row} J.S. Rowlinson, {\it Liquids and Liquid Mixtures}, Butterworths,
London, 1971.
%
\bibitem{McH-K} M.A. McHugh and V.J. Krukonis, {\it Supercritical extraction},
Butterworths, London, 1985.
%
\bibitem{ww} E. C. Zeeman, {\it Catastrophe Theory, Selected Papers 1972-1977},
Reading, Addison-Wesley, 1977.
%
\bibitem{lander} Th. Br\"ocker and L. Lander, {\it Differentiable germs and
catastrophes}, Cambridge University Press, 1975.
%
\bibitem{LNM} A.E.R. Woodcock and T. Poston, {\em A geometrical study of 
elementary Catastrophes}, Lecture Notes in Mathematics {\bf 373}, 
Springer-Verlag 1974
%
\bibitem{gmm} J. Gaite, J. Margalef-Roig and S. Miret-Art\'es, in 
preparation.



\end{thebibliography}
\end{document}